\documentclass[a4paper,fleqn]{cas-dc}

\usepackage[square,numbers]{natbib}

\usepackage{babel}

\usepackage{microtype}
\usepackage{nameref}
\usepackage{scalerel}
\usepackage{lipsum,adjustbox}
\usepackage{graphicx}
\usepackage{caption}
\usepackage{subcaption}
\usepackage{tikz}
\usepackage{tkz-graph}  
\usepackage{arydshln}
\usepackage{hyperref}
\usepackage{enumitem}
\usepackage{amssymb}
\usepackage{pifont}% http://ctan.org/pkg/pifont

\usetikzlibrary{graphs}
\usetikzlibrary{shapes.geometric}% 
\usetikzlibrary{positioning}
\usetikzlibrary{arrows.meta}
\newcommand{\cmark}{\ding{51}}
\newcommand{\xmark}{\ding{55}}

\begin{document}
\let\WriteBookmarks\relax
\def\floatpagepagefraction{1}
\def\textpagefraction{.001}

\shorttitle{Virtual Harassment, Real Understanding: Using a Serious Game and Bayesian Networks to Study Cyberbullying}

% \shortauthors{J. Pérez \textit{et~al}.}
\shortauthors{J. Pérez}

\title [mode = title]{Virtual Harassment, Real Understanding: Using a Serious Game and Bayesian Networks to Study Cyberbullying}               %Generating Synthetic Data in a Serious Game to Fight Cyberbullying


\author[1]{Jaime Pérez}[orcid=0000-0001-7511-2910]
\cormark[1]
\ead{jperezs@comillas.edu}

\author[1,2]{Mario Castro}[orcid=0000-0003-3288-6144]

\author[3]{Edmond Awad}[orcid=0000-0001-7272-7186]

\author[1]{Gregorio López}[orcid=0000-0001-9954-3504]

\affiliation[1]{organization={Institute for Research in Technology (IIT), ICAI Engineering School, Universidad Pontificia Comillas},
    %addressline={Radarweg 29}, 
    city={Madrid},
    % citysep={}, % Uncomment if no comma needed between city and postcode
    postcode={28015}, 
    country={Spain}}
\affiliation[2]{organization={Grupo Interdisciplinar de Sistemas Complejos (GISC), Department of Mathematics, Universidad Carlos III},
    %addressline={Radarweg 29}, 
    city={Madrid},
    % citysep={}, % Uncomment if no comma needed between city and postcode
    postcode={28911}, 
    country={Spain}
    }

\affiliation[3]{organization={Department of Economics, University of Exeter},
    %addressline={Radarweg 29}, 
    city={Exeter},
    % citysep={}, % Uncomment if no comma needed between city and postcode
    postcode={EX4 4PU}, 
    country={United Kingdom}}

\cortext[cor1]{Corresponding author}

%----------ABSTRACT---------
\begin{abstract}
Cyberbullying among minors is a pressing concern in our digital society, necessitating effective prevention and intervention strategies. Traditional data collection methods often intrude on privacy and yield limited insights. This study explores an innovative approach, employing a serious game - designed with purposes beyond entertainment - as a non-intrusive tool for data collection and education.
In contrast to traditional correlation-based analyses, we propose a causality-based approach using Bayesian Networks to unravel complex relationships in the collected data and quantify result uncertainties. This robust analytical tool yields interpretable outcomes, enhances transparency in assumptions, and fosters open scientific discourse.
Preliminary pilot studies with the serious game show promising results, surpassing the informative capacity of traditional demographic and psychological questionnaires, suggesting its potential as an alternative methodology.
Additionally, we demonstrate how our approach facilitates the examination of risk profiles and the identification of intervention strategies to mitigate this cybercrime. We also address research limitations and potential enhancements, considering the noise and variability of data in social studies and video games.
This research advances our understanding of cyberbullying and showcase the potential of serious games and causality-based approaches in studying complex social issues.

\end{abstract}

% \includegraphics{figs/grabs.pdf}

%----------HIGHLIGHTS---------
\begin{highlights}
\item This paper innovatively utilizes serious games as a non-invasive tool for data collection and education to study cyberbullying among minors.

\item We advocate for a causality-based approach using Bayesian Networks to analyze the collected data, providing more robust and interpretable results compared to traditional correlation or ML-based methods.

\item The results suggest that serious games offer the potential to provide more valuable information than demographic data, enhancing our understanding of minors' online behavior in the context of cyberbullying.
\end{highlights}

%----------KEYWORDS---------

\maketitle

%----------INTRODUCTION---------
\section{Introduction}
\label{intro}

The rise of the digital age has led to an increase in children's Internet use. In 2017, minors under 18 accounted for almost one-third of all Internet users worldwide, as reported by UNICEF's "Children in a Digital World" study \cite{UNICEFchildren2017}. With the COVID-19 pandemic, this trend has only intensified, with children spending more time online at a younger age. Unfortunately, unregulated access to the Internet also exposes children to new dangers. Those who were already vulnerable are now at an even greater risk. For example, approximately 10\% of European children are cyberbullied every month \cite{EUkidsOnline2020}, and nearly half have experienced a cyberbullying incident at least once \cite{ChidrenRisk_Covid2021}.

Research on cyberbullying in traditional social science often involves prevention strategies that are difficult to scale and invasive methods like questionnaires, which are unappealing to minors. There is a cultural gap between the language or format of the surveys and how minors perceive them. Questionnaires are often seen as assessments with grades attached, contributing to this negative perception. This barrier demands novel approaches aligned with their cultural background.

Alternatively,  Serious Games (SG) are an appealing alternative approach as they can be used as an educational and research tool to promote a preventive approach. SGs are designed not just for entertainment but also to teach new skills, convey values, and raise awareness \cite{abt1987serious} (e.g., the H2020 RAYUELA project \cite{RAYUELA}). Despite their educational purpose, SGs remain appealing and entertaining. The RAYUELA project aims to leverage the natural appeal of video games to gather data and educate minors. Players are immersed in an interactive visual adventure, making decisions related to cybercrime. Data will be collected and analyzed using modern data science methods to understand better the factors influencing risky online behavior in a friendly, safe, and non-invasive manner.

Although SGs are widely recognized for their educational value, their potential as social research tools remains largely unexplored \cite{gamesScience2023, PerezSeriousGames}. However, there have been notable recent examples of their use. For instance, a video game experiment found that the complexity of the city in which a child resides affects their future navigation skills \cite{Navigation2022}. Similarly, an English grammar game named "\textit{Which English?}" revealed that there is a "critical period" for learning a second language that extends into adolescence \cite{English2018}. Or "\textit{The Moral Machine}" experiment, a dilemma-based game with millions of players that explored the moral values of different societies and how they vary across countries \cite{awad_moral_2018}.

One of the unexploited benefits of SGs is their ability to gather, non-intrusively, data from the player's decisions. However, navigating data analysis is a significant challenge within cyberbullying research. Traditional methodologies often employ basic statistical techniques that seek correlations, overlook the potential biases introduced by confounders or colliders variables, or focus on p-value hypothesis testing. Alternatively, some researchers embrace the promise of the impressive predictive capabilities of machine learning algorithms~\cite{rahal2022rise}, which, while powerful, might build their predictions on unrealistic patterns or spurious relationships that are neither easily interpretable nor explainable. 

In this paper, we suggest a different approach to analyzing cyberbullying. Instead of relying on traditional methods, we propose a causality-based perspective better suited to handling the sensitive nature of the issue~\cite{pearl2009causality}. This approach is especially useful when working with limited or unreliable data and is essential when incorporating expert knowledge into statistical models~\cite{Pearl_Do_Calculus}. To achieve this, we use Bayesian Networks, a probabilistic graphical model that depicts the causal relationships between variables clearly and understandably. 
%By taking this approach, we hope to gain a more precise and nuanced understanding of cyberbullying, which can help inform more effective prevention and intervention strategies.

In this context, three questions arise naturally: (i) What are the risk factors that \emph{cause} an increased risk of being a cyberbullying offender?; (ii) can we combine attitudes in an SG with background information to \emph{profile} risky attitudes; and (iii) can we use a SG as an alternative (or complement) of a validated questionnaire?

The main objective of this article is to contribute to a richer, more comprehensive understanding of the cyberbullying phenomenon, which can help inform more effective prevention and intervention strategies. This under\-standing is derived from experimental pilot studies involving minors using a SG and a causal interpretation of the data using expert knowledge pondered with actual data. The causal approach reduces the biases introduced by confounding while helping to design effective interventions.

% using an innovative tool: a serious game. %In this context, serious games emerge not merely as platforms for entertainment or education but as powerful instruments for social research, fostering engagement, and providing insights into the behavior and decision-making processes of the participants.
%We employ causality-based methods for data analysis (specifically, Bayesian networks), aiming to yield more robust and reliable conclusions than those emerging from more basic statistical approaches. 
% Ultimately, this work aims to highlight the potential of serious games as social research tools and to demonstrate the effectiveness of causal thinking in revealing nuances about complex social phenomena such as cyberbullying. Through this approach, we aspire to provide deeper insights into the factors and dynamics that drive cyberbullying, contributing to developing more effective prevention and intervention strategies.

%----------METHODOLOGY---------
\section{Methodology}
This section briefly describes the SG developed and outlines the data collection procedure. It emphasizes the variables collected during pre- and post-game sessions. Lastly, the strategies employed to analyze the data with a causality mindset are elaborated and justified.

\subsection{Serious Game Description}
As outlined in the previous section, the primary focus of the RAYUELA project \cite{RAYUELA} is to study in depth the drivers and human factors contributing to specific types of cybercrime affecting minors. This goal is achieved through a unique approach that leverages gaming, providing a platform for learning and modeling behaviors in an engaging and non-invasive way.

\begin{figure}[]
  \center
  \caption{Screen capture from the RAYUELA SG.}
  \includegraphics[width=0.48\textwidth,keepaspectratio]{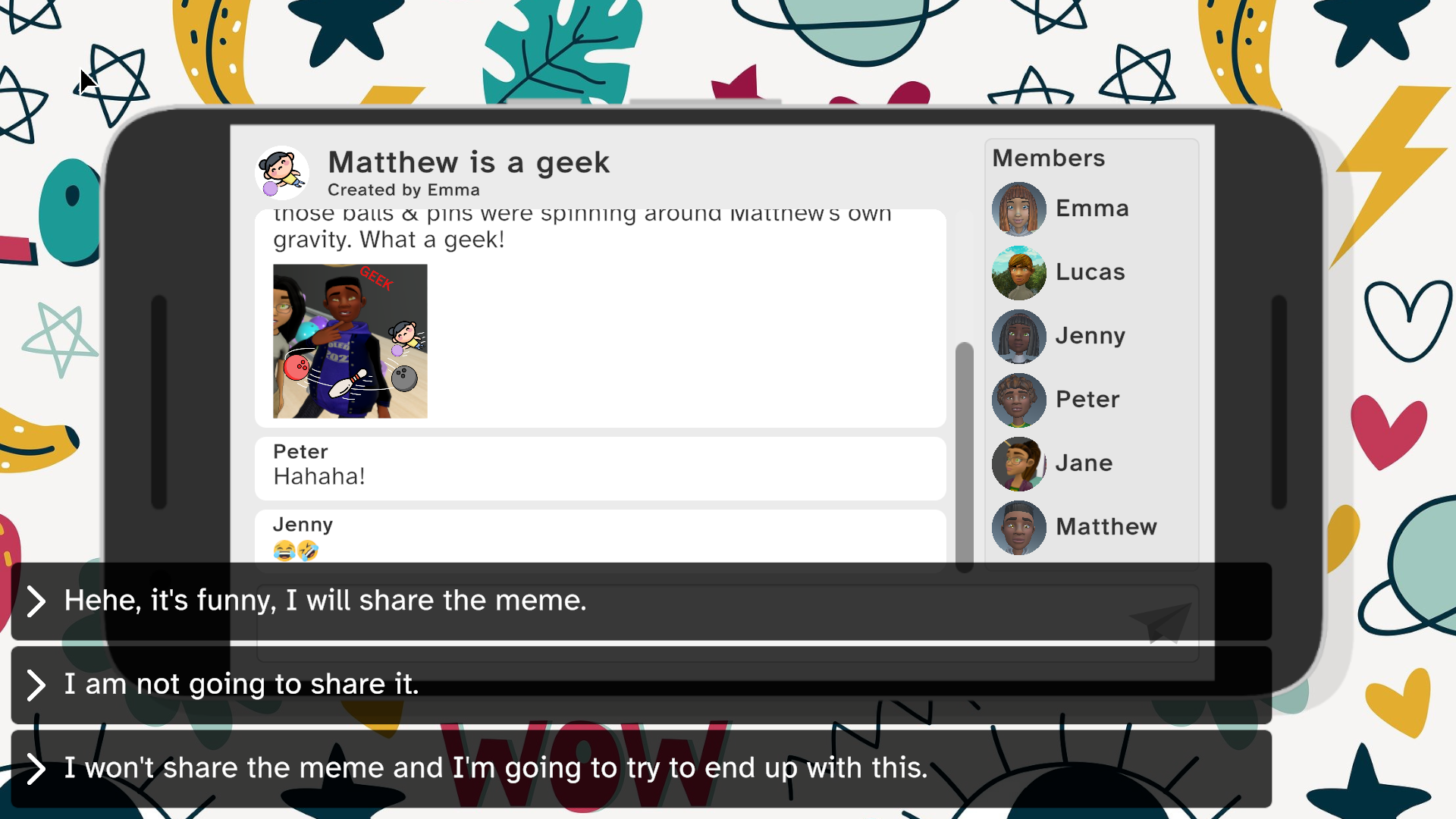}
  \label{fig:Screen}
\end{figure}

The SG, developed by the Tecnalia\footnote{\url{https://www.tecnalia.com/}} team (part of the RAYUELA consortium), is a point\&click 3D graphic interactive narrative adventure. Players make decisions that shape the narrative's progression and outcome. The game is set in a high school, presenting scenarios involving cybercrimes affecting young individuals, such as fake news, cyberbullying, and online grooming. Fig.~\ref{fig:Screen} depicts a screen capture of the SG.

The player's decisions in the game are meticulously designed and discussed by the RAYUELA team to analyze player patterns and identify those most susceptible to committing or experiencing specific cybercrimes. The game comprises several adventures, each of which addresses a cybercrime or the same cybercrime from a different perspective. Demographic and psychological questionnaires are also administered during the gaming session to validate the researchers' hypotheses. This allows the measurement of certain variables or potential cybercrime drivers in-game and out-of-game. Table \ref{tab:Variables} presents the cyberbullying indicator or driver variables identified by the RAYUELA team through literature review, interviews, and focus groups, along with indications of in-game or out-of-game measurement.

% \usepackage{graphicx}
\begin{table}[]
\caption{Indicator and driver variables of cyberbullying identified, illustrating whether each variable was measured in-game or out-of-game.}
\label{tab:Variables}
\resizebox{0.49\textwidth}{!}{%
\begin{tabular}{llcc}
\toprule
\textbf{Type} & \textbf{Indicator/driver} & \textbf{\begin{tabular}[c]{@{}c@{}}Measured\\ in-game\end{tabular}} & \textbf{\begin{tabular}[c]{@{}c@{}}Measured\\ out-of-game\end{tabular}} \\
\midrule
Environmental & \begin{tabular}[c]{@{}l@{}}Isolation/lack of\\ social support\end{tabular} & \cmark & \cmark \\
 & Family communication & \cmark & \cmark \\
 & Socio-economic status & \xmark & \xmark \\
 & School type & \xmark & \xmark \\\hline
Personal & Previous CB Victimization & \cmark & \cmark \\
 & Low self-esteem & \cmark & \cmark \\
 & \begin{tabular}[c]{@{}l@{}}Difficulty in making\\ friends face to face\end{tabular} & \cmark & \xmark \\
 & Poor mental health & \xmark & \xmark \\
 & Empathy & \cmark & \cmark \\
 & Age & \xmark & \cmark \\
 & Gender & \xmark & \cmark \\
 & Sexual orientation & \xmark & \cmark \\
 & Migratory background & \xmark & \cmark \\\midrule
Technological & \begin{tabular}[c]{@{}l@{}}Public profile on social\\ networks and publishing\\ a lot of information\end{tabular} & \cmark & \xmark \\
 & Time spent online & \cmark & \cmark \\
 & \begin{tabular}[c]{@{}l@{}}Weak passwords and\\ share them\end{tabular} & \cmark & \xmark \\
\bottomrule 

\end{tabular}%
}
\end{table}

\subsection{Data Acquisition}
The methodology involved gathering data from minors through our SG, with collaborations from European schools and institutes in Greece, Belgium, and Spain. We received 224 responses from students aged 12 to 16 (Mean=13.6, SD=1.37), with 53\% identifying as males, 44\% as females, and 1\% as non-binary. Before starting the game, all participants registered and completed demographic and psychological questionnaires, constituting the pre-game data collection phase. The following variables were obtained during this phase:

\begin{itemize}
    \item Demographic Variables: Age, gender, sexual orientation, migratory background, and daily hours spent on the Internet. These variables were considered to understand the diverse demographic background of the participating minors, along with a measure of their relationship with technology.
    \item Psychological Variables: Empathy, social and family support, and self-esteem. These variables were chosen to provide a baseline for understanding the minors' emotional, social, and psychological state before playing the game. They were extracted from the following validated questionnaires: the \emph{Rosen\-berg self-esteem scale} \cite{rosenberg1965rosenberg} and \emph{The Multidimensional Scale of Perceived Social} Support \cite{SocialSupport}.
\end{itemize}

The project's cyberbullying experts meticulously chose each demographic and psychological variable, aligning with prior research suggesting these factors substantially impact the susceptibility and response to cyberbullying. Table \ref{tab:EDA} provides variable values and their percentage of occurrences (marginal probability).

% Regarding the data collected through the serious game, the following variables were obtained: Answers to game decisions/questions, response times, and a final question about the differences between the player behavior in the game and the real world ("Have you played as you would behave in the real world?"). This final question was proposed as a calibration method for the answers given in the serious game, as some players may react differently than they usually do when they are online. Another indicator of the «honesty» of the players in their game responses may be the gameplay reaction times. That is, we can reasonably assume that answers with extremely short reaction times are random choices of the player.

The gameplay data encompasses three key elements: (i) players' decisions at each game dilemma, (ii) response times to different situations, and (iii) a post-game questionnaire evaluating the alignment between in-game choices and real-world behavior ("Did you play as you would behave in the real world?"). This \emph{honesty} question acts as calibration and control, as the game format might encourage adventurous choices for exploration. 
%The Bayesian framework accommodates contradictory playing styles as both approaches to the game are informative about the players' background and demographics. 
Additionally, short reaction times may indicate random choices. 

After the game, minors completed a questionnaire about their past cyberbullying experiences, serving as ground truth for the analyses. The validated questionnaire used was the \emph{European Cyberbullying Intervention Project Questionnaire}~\cite{CB_Questionnaire}.
 % A more exhaustive exploratory data analysis is provided in Appendix \ref{appendixEDA}.

\begin{table}[]
\caption{Model variables from registration (demographics) and validated questionnaire~\cite{CB_Questionnaire}. The possible response values of each variable and its sample marginal probability (i.e., percentage of observation) are shown.}
\label{tab:EDA}
\resizebox{0.49\textwidth}{!}{%
\begin{tabular}{lll}
\toprule
\textbf{Variable} & \textbf{Response Values} & \textbf{Marginal Probability} \\
\midrule
Gender & Male & 62.9\% \\
 & Female & 35.1\% \\
 & Non Binary & 1\% \\
Age & 12 & 18.8\% \\
 & 13 & 4.4\% \\
 & 14 & 26.8\% \\
 & 15 & 33\% \\
 & 16 & 17\% \\
Sexual Orientation & Heterosexual & 55.3\% \\
 & Non Heterosexual & 5.4\% \\
Migratory Background & No & 71.4\% \\
 & \begin{tabular}[c]{@{}l@{}}My parents were born\\ in another country\end{tabular} & 8.6\% \\
 & \begin{tabular}[c]{@{}l@{}}I was born in another\\ country\end{tabular} & 20\% \\
Self-Esteem & Low & 37.5\% \\
 & Medium & 41.5\% \\
 & High & 21\% \\
Social Support & Low & 3.6\% \\
 & Medium & 33.5\% \\
 & High & 62.9\% \\
Family Support & Low & 7.6\% \\
 & Medium & 24.5\% \\
 & High & 67.9\% \\
Daily Hours of Internet & Less than 1 h & 8.9\% \\
 & 1-2 h & 18.7\% \\
 & 2-3 h & 21.4\% \\
 & 3-4 h & 33\% \\
 & More than 4 h & 15.6\% \\
Empathy & Low & 45.8\% \\
 & High & 54.2\% \\
\bottomrule
\end{tabular}%
}
\end{table}

\subsection{Causality-based Methodology: Bayesian Networks}
Our main objective in this research is to gain a deeper understanding of cyberbullying and the interconnectedness of its variables rather than focusing solely on algorithmic predictive power. To achieve this, we have implemented a causality-based methodology for thorough analysis.

Causality-based approaches are particularly useful when studying sensitive topics like cyberbullying, as they reveal the underlying processes rather than just the ability to make predictions. Understanding causal relationships is critical in identifying intervention points to optimize prevention and detection strategies, ultimately mitigating the negative effects of cyberbullying.

We have utilized Bayesian Networks (BNs), a probabilistic graphical model that captures statistical and causal relationships between variables. BNs use directed acyclic graphs (DAG) to represent dependencies efficiently, leading to improved interpretability and transparency of the assumptions adopted by the investigators. Additionally, these networks can effectively detect confounding factors and minimize potential bias. They can also handle counterfactual analyses and simulate interventions, answering "what-if" questions based on available evidence.

\begin{figure*}[]
  \center
  \caption{Structure of the BN proposed by the cyberbullying experts from the RAYUELA project. The variable of interest (\emph{Previous CB Offending}) is highlighted in blue.}
  \includegraphics[width=0.66\textwidth,keepaspectratio]{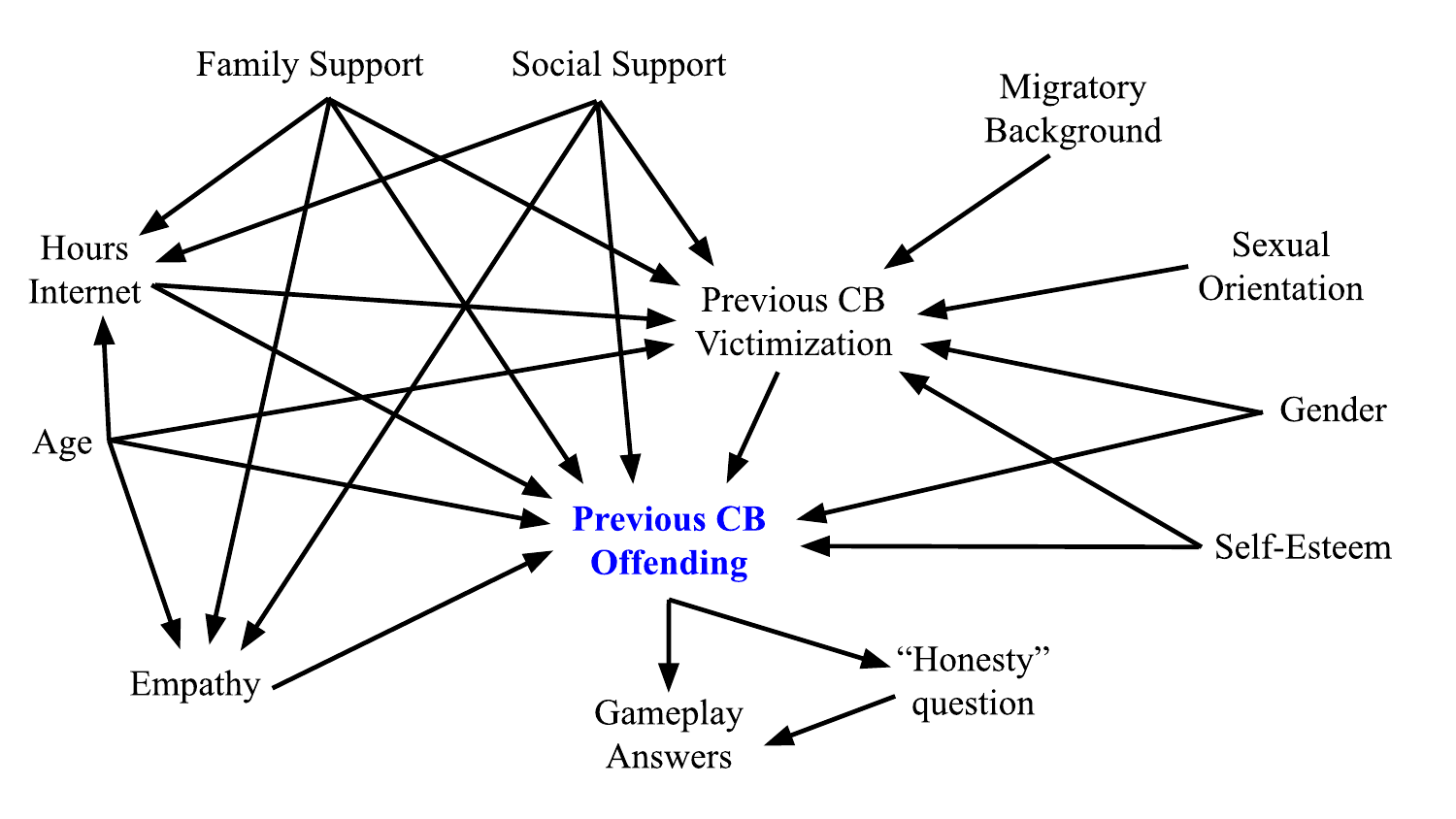}
  \label{fig:DAG}
\end{figure*}

Constructing a coherent BN structure that aligns with reality is crucial. Although data-driven algorithms exist for inferring BN structures, incorporating expert knowledge, especially in data-limited scenarios, is vital. Thus, we used a causal structure provided by cyberbullying experts for the model's validity and accuracy. Fig.~\ref{fig:DAG} displays the BN architecture proposed by these experts. 

Once the BN structure has been validated, we train its parameters (i.e., the conditional probability tables) using the available data. We then perform three analyses to \emph{interrogate} the model, as described in Sec.~\ref{sec:results}.

%----------RESULTS---------
\section{Results}
\label{sec:results}
This section outlines the experimental procedures for extracting valuable insights from the available data. Several analyses and experiments can be conducted based on the constructed BN structure, such as exploring causal relationships, simulating interventions, assessing predictive capabilities, and other related investigations. However, our paper primarily concentrates on the analyses that offer the most significant contributions to comprehending cyberbullying as a phenomenon. In particular, we articulate our research around three scientific questions:% we address the following analyses:

\begin{enumerate}
    \item \textbf{Which variables are more strongly associated with the risk of being a CB offender? Quantifying Arrow Strength:} 
    
    Measure the strength of influence of each arrow on the variable of interest (i.e., CB offending). They allow us to understand each variable's individual contribution to the model output.
    % \item \textbf{What combinations of variables allow to build significant risk profiles? Multi-Factor Analysis:} 
    
    \item \textbf{What combinations of variables make up the risk profiles identified? Multi-Factor Analysis:}
    
    Evaluate combinations of multiple variables and their collective impact on the variable of interest (i.e., CB offending). This analysis aims to identify which combinations of characteristics are most common in risk profiles.
    
    \item \textbf{Can a Serious Game be used as a scientific instrument to quantify CB risk? Latent Variable Model:} 
    
    Assess inferences that could be drawn without labeled data (i.e., without the cyberaggression/victimization questionnaire). This analysis poses a validation step to justify using the SG as a social research tool.
    % {\color{red}
    % \item \textbf{What impacts the conclusions of our prior assump\-tions about CB offending risk prevalence? Sensitivity analysis:} 
    
    % Finally, we conduct a sensitivity analysis of the prior used in the prevalence of previous CB Offending to assess the robustness of the data analysis to pre-existing epidemiological uncertainty.
    % }
\end{enumerate}

\subsection{Which variables are more strongly associated with the risk of being a CB offender? Quantifying Arrow Strength}
\label{Analysis1}
This initial analysis, commonly referred to as "strength of influence," aims to address the question: \emph{How strong is the causal impact of a cause on its direct effect?} In simpler terms, we will quantify the strength of influence of certain variables on a variable of interest. In this case, the variable of interest is \textit{Previous CB offending} (i.e., engaging in cyberbullying acts in the past). This analysis provides insights into the most relevant variables influencing the model results, considering them individually. That is, without considering combinations of variables.

The method used for this analysis is based on Koiter's work \cite{koiter2006visualizing}. It primarily involves measuring the similarity between multiple probability distributions of the variable of interest based on the states of the parent nodes. The chosen metric to calculate this similarity is the Jensen-Shannon distance \cite{JSD}, normalized between 0 and 1 to ensure interpretability.

\begin{figure*}[]
  \center
  \caption{Analysis 1: The strength of influence of each variable on the output (\emph{Previous CB offending}) in the BN is analyzed. This involves measuring the distance between the probability distributions of each variable while marginalizing \emph{Previous CB offending} in its two possible values. The normalized Jensen-Shannon distance (0-1) serves as the distance measure. "\emph{Adventure 1 - Question 1 - Photo Sharing}" acts as a control variable with random answers, making variables with similar or lower metrics irrelevant for practical purposes.}
  \includegraphics[width=0.85\textwidth,keepaspectratio]{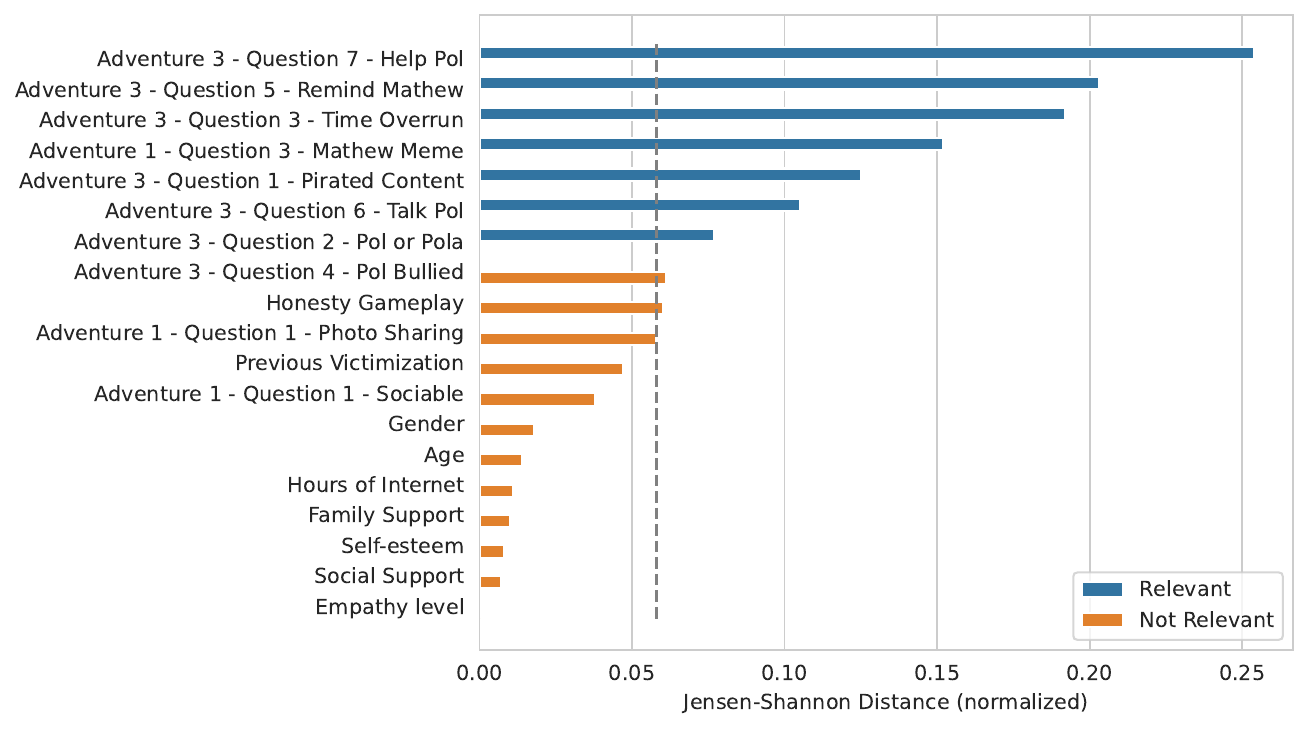}
  \label{fig:Analysis1_Aggression}
\end{figure*}

Fig.~\ref{fig:Analysis1_Aggression} presents the analysis outcomes concerning the variable of interest, \textit{Previous CB offending}. The variable \textit{A1\_1\_PhotoSharing} acts as a control, generating random responses from players. Hence, variables with similar or lower metrics than \textit{A1\_1\_PhotoSharing} are considered irrelevant for practical purposes. The results indicate that the relevant variables are solely derived from SG responses, assuming the accuracy of the constructed BN and analyzing each variable's individual influence. This implies that the SG could be a reliable predictor for the variable of interest, \textit{Previous CB offending}. On the other hand, the demographic variables, when examined individually, exhibit minimal significance.
Nevertheless, these demographic variables are expected to exhibit stronger associations with larger numbers of participants, yet this underscores the usefulness of the game and BN for small sample sizes.

\begin{figure}[]
  \center
  \caption{Analysis 1: Conditional probabilities of the affirmative case of the output (\emph{Previous CB offending}) marginalized on the possible values of the variable "\emph{Adventure 3 - Question 7 - Help Pol}". According to the results of analysis 1, this variable is the most influential on the output, considering the individual level of each variable in the model. This means that marginalizing the variable's possible values changes the output's conditional probability (\emph{Previous CB offending}) significantly.}
  \includegraphics[width=0.48\textwidth,keepaspectratio]{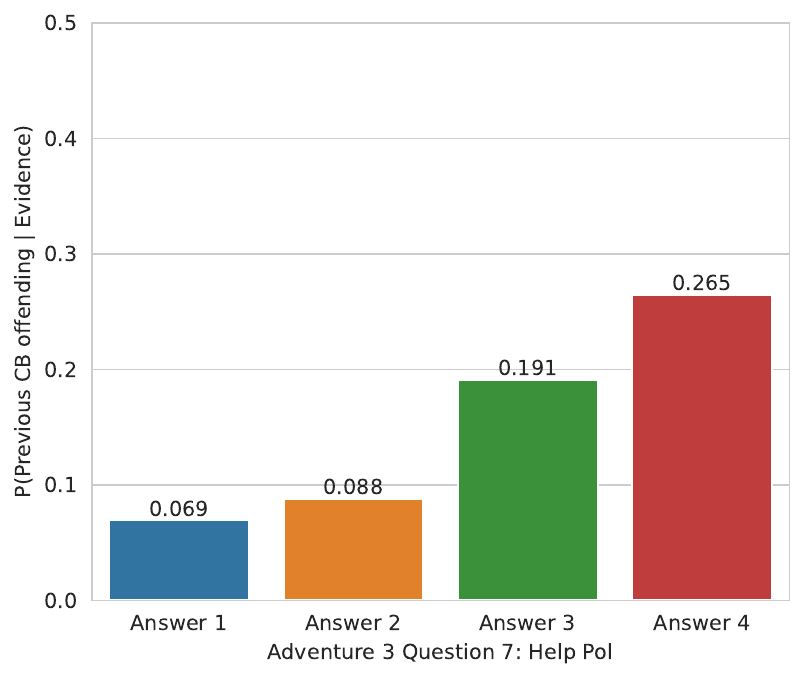}
  \label{fig:Analysis1_Aggression_CP}
\end{figure}

Fig.~\ref{fig:Analysis1_Aggression_CP} elucidates the significance of these metrics. It presents the conditional probabilities of \textit{Previous CB offending} based on all potential values for the variable with the strongest influence: \textit{Adventure 3 Question 7: Help Pol}. This figure demonstrates that players choosing answer 4 have a 26.5\% likelihood of engaging in cyber-aggression in the past, independent of other variables. Conversely, those selecting answer 1 have a mere 7\% chance.

\subsection{What combinations of variables make up the risk profiles identified? Multi-Factor Analysis}
\label{Analysis2}
This analysis examines the importance of the variables in the model. However, unlike in the fisrt analysis (\ref{Analysis1}), we are now interested in finding combinations of variables (i.e., multi-factor) that significantly change the conditional probability of the variable of interest, in this case, \textit{Previous CB offending}. In addition, we will use this analysis to compare the relevance of variables coming from the gameplay with those from demographic variables or psychological questionnaires.

The variables that do not come from the gameplay are encompassed in the term "profiling." Player profiling consists of analysis or categorization using only static variables that are not (necessarily) directly related to gameplay \cite{yannakakis2018artificial}. Those are demographic variables (age, gender, sexual orientation, migratory background, daily hours of Internet use) and those collected through psychological questionnaires (social support, family support, self-esteem, previous CB victimization, and previous CB offending).

The method used in this analysis consists of inserting multiple observations into the BN and recording the probabilities that have been updated using Bayes' formula \cite{Jensen2007}. To do this, we will brute force all combinations of values from 1 to 10 fixed evidence for both cases (game questions and profiling). For example, 1 fixed evidence could be \emph{Age=14} or \emph{Adventure 1 Question 3: Mathew Meme=Answer 2}.

\begin{figure}[]
  \center
  \caption{Analysis 2: By performing a multi-factor analysis, we can find the combinations of variables that cause a greater increase in the conditional probability of the outcome (\emph{Previous CB offending}). The following figure is obtained by finding the maximum conditional probability obtained by setting different numbers of combinations of evidence. This is done, on the one hand, for the variables obtained through the game questions and, on the other hand, for the profiling variables. The figure also shows the conditional probabilities corresponding to the relevant thresholds of the Bayes Factor according to Jeffreys' criterion \cite{jeffreys1998theory} calculated in Equation \ref{eq:odds}.}
  \includegraphics[width=0.48\textwidth,keepaspectratio]{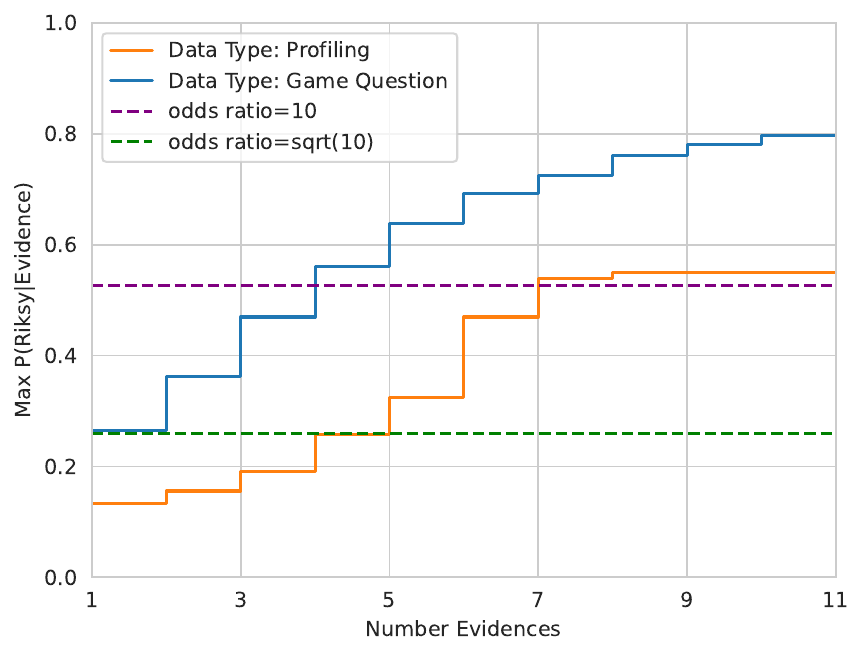}
  \label{fig:Analysis2}
\end{figure}

Fig.~\ref{fig:Analysis2} shows the maximum conditional probability of a positive response to the \textit{Previous CB offending} variable as a function of the amount of evidence inserted in the BN. Results are presented for both data sources (game questions and profiling). Also shown are two lines marking significant values of the conditional probability of the variable of interest compared to the prior probability distribution, which was set to 0.1 before training the BN parameters, with an effective sample size of 2 (soft prior). Using the Jeffreys scale \cite{jeffreys1998theory} for comparing odds ratios (i.e., Bayes Factor), a ratio between $10^{1/2}$ and $10$ is interpreted as a substantial difference. A ratio between $10$ and $10^{3/2}$ is a strong difference. In our case, with 0.1 prior probability, this would occur with posterior probabilities of $\sim0.26$ and $\sim0.53$, respectively (Equation \ref{eq:odds}). Although, it is essential to remember that as the number of fixed pieces of evidence increases, the number of players who meet these criteria will decrease.

This analysis confirms the conclusion drawn in Analysis 1: The variables obtained through gameplay are more effective in distinguishing players who have previously engaged in cyberbullying offending from those who have not.

\begin{equation}
    \begin{aligned}
        BF &= \frac{\text{prior odds ratio}}{\text{posterior odds ratio}} = \frac{0.9/0.1}{(1-X)/X} \\
        BF &= 10^{1/2} \text{ (Substantial evidence)} \Rightarrow X \approx 0.26 \\ 
        BF &= 10 \text{ (Strong evidence)} \Rightarrow X \approx 0.53 \\
    \end{aligned}
    \label{eq:odds}
\end{equation}

\begin{figure*}[]
  \center
  \caption{Analysis 2: Through a multi-factor analysis, we identify "risky" combinations of variables (profiles) for cyberbullying. The figure displays the count of occurrences for each variable value in the identified risk profiles. We focus on profiles with exactly five variables, as they yield significant odds ratios (i.e., Bayes Factor) following Jeffreys' criterion \cite{jeffreys1998theory}. Notably, \emph{Previous CB Victimization = True} appears in over 80\% of the risky profiles considered.}
  \includegraphics[width=\textwidth,keepaspectratio]{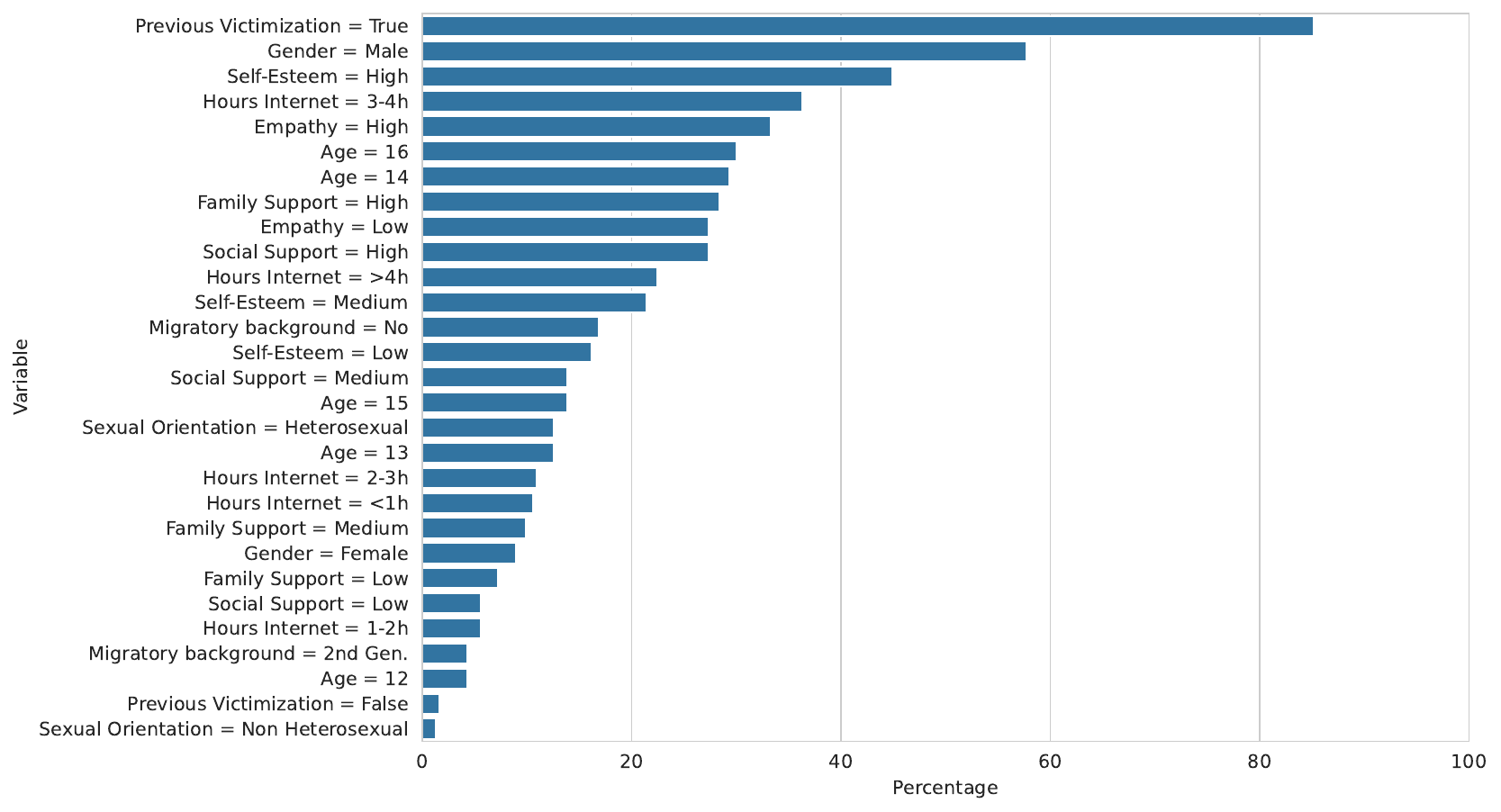}
  \label{fig:Profiling}
\end{figure*}

After examining the five pieces of evidence, we found that the profiling variables exceeded the initial threshold of around 0.26. Upon further analysis, Fig.~\ref{fig:Profiling} displays the frequency of observations for common shared characteristics among risk profiles. A \emph{risk profile} is defined as having a posterior probability (\textit{Previous CB offending=Yes}) of 0.26 or greater. The top shared profiling characteristics among these risk profiles are: \textit{previous CB victimization=True, gender=Male}, and \textit{self-esteem=High}. 

% \begin{table}[width=.99\linewidth,cols=3,pos=h]
% \caption{XXXXXXXXXXXXXXX}
% \label{tab:Analysis2_Aggressor}
% \begin{tabular*}{\tblwidth}{@{} LLL@{} }
% \toprule
% Nº Evidences & \begin{tabular}[l]{@{}l@{}}Highest \\ $P(\text{\emph{aggression}}=$\text{\emph{Yes}}|\text{\emph{Evidences}}) \end{tabular} & Odds ratios \\
% \midrule
% 1 & 0.132 & 1.368 \\
% 2 & 0.155 & 1.651 \\
% 3 & 0.191 & 2.125 \\
% 4 & 0.258 & 3.129 \\ \hdashline
% 5 & 0.324 & 4.313 \\
% 6 & 0.469 & 7.95 \\ \hdashline
% 7 & 0.538 & 10.48 \\
% 8 & 0.55  & 11 \\
% 9 & 0.55  & 11 \\
% 10 & 0.55 & 11 \\
% \bottomrule
% \end{tabular*}
% \end{table}

\subsection{Can a Serious Game be used as a scientific instrument to quantify CB risk? Latent Variable Model}
This analysis aims to answer the following question: What conclusions would we have drawn if we had not labeled data? That is, if the players had not completed the cyber-victimization/aggression questionnaire.

For this purpose, we will perform the exact experiment from the first analysis (\ref{Analysis1}) for a BN model with a latent or unobserved variable (\emph{Previous CB offending}). In this case, the BN training algorithm will perform a type of Bayesian clustering. That is, it will divide the players into two groups based on the data from the rest of the nodes in the network and the prior probability distribution. Once this second ranking is obtained, we will make a quantitative comparison with the ranking obtained in Analysis 1 using Spearman's correlation \cite{Spearman}. If this correlation is high (close to 1), it will suggest that the rankings obtained in both cases are very similar, and therefore, similar conclusions will be drawn from the total and latent models.

\begin{figure*}[]
  \center
  \caption{Analysis 3: A model identical to the one used in the previous analyses was developed, but where the variable of interest is a latent state (i.e., we have no labeled data). We then performed the same experiment as in the first analysis (\ref{Analysis1}), where we ranked the importance of each variable in affecting the conditional probabilities of the latent outcome (\emph{Previous CB offending}). Again, the normalized Jensen-Shannon distance (0-1) is used as the distance measure. And "\emph{Adventure 1 - Question 1 - Photo Sharing}" is a control variable in which players give random answers. Therefore, variables with similar or lower metrics than this can be considered irrelevant for practical purposes.}
  \includegraphics[width=0.85\textwidth,keepaspectratio]{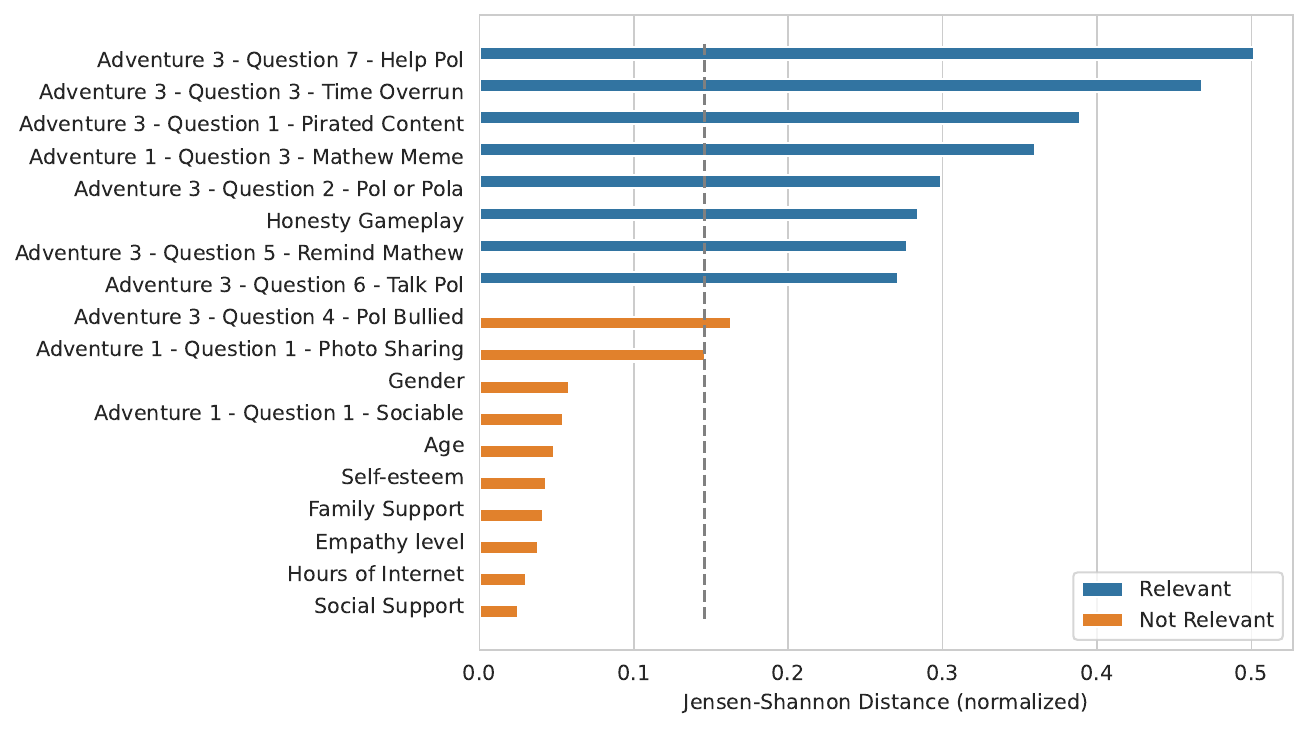}
  \label{fig:Analysis3}
\end{figure*}

Fig.~\ref{fig:Analysis3} analysis results on the latent variable of interest \textit{Previous CB offending}. As in Analysis 1, the variable \textit{A1\_1\_PhotoSharing} is a control variable. Therefore, variables with similar or lower metrics than \textit{A1\_1\_PhotoSharing} can be considered irrelevant for practical purposes. The results show again that the only relevant variables are those obtained through the SG responses, analyzing the strength of influence of each variable individually. The Spearman correlation for both rankings is 0.923 (with a p-value of 4.57e-08).

These results seem to indicate that even if we did not have labeled data (i.e., the cyber-victimization/aggression questionnaire), we would have been able to draw very similar conclusions: When looking at them individually, the variables obtained through the gameplay appear to be much more relevant in distinguishing between players who have engaged in cyberbullying aggression in the past and those who have not.

% \subsection{Analysis 4: Sensitivity to prior on prevalence}
%----------DISCUSSION---------
\section{Discussion and Limitations}
Our research provides intriguing evidence that data derived from SGs can offer deeper insights into cyberbullying behaviors than traditional profiling variables such as demographic data and psychological assessments. These data analyzed both individually (\ref{Analysis1}) and in multi-factor analysis (\ref{Analysis2}) point to the efficacy of serious gaming as a social research tool.

This novel revelation underscores the potential of SGs to provide real-time, experience-based observations on sensitive and multifaceted social phenomena such as cyberbullying, which standard profiling approaches might inadequately address. Moreover, such tools could provide a fun and non-invasive way to obtain information from hard-to-reach populations, such as minors.

Despite these promising revelations, we must be cautious when interpreting our results. We must bear in mind the limitations of our research, particularly the limited sample size and the small geographic group of participants from only three European countries. We are, moreover, considering that social science and video game data can be particularly noisy. A more extensive and diverse data set would contribute to the robustness and generalizability of our conclusions.

Furthermore, the robustness of our (causality-based) BN analyses depends heavily on the reliability of the structure devised by the experts. Any modification of this network structure could invariably influence the results, a contingency to be considered when interpreting our conclusions.

Despite these caveats, our findings are encouraging, indicating that SGs may serve as valuable research tools. BNs demonstrated their utility in our study, effectively detecting confounders and collider variables, thereby reducing bias. In addition, they promote transparency by making the underlying assumptions of the researcher explicit, contributing to more accessible and comprehensible science.

Our future efforts will be directed towards augmenting the data collection, thereby amplifying the sample size for further validation or refutation of our findings. The application of Explainable Machine Learning techniques could be explored for comparative analyses. Though their prediction-centric focus may offer a different rigor for dealing with sensitive issues, the comparative results might provide additional valuable insights.

Finally, while BNs underscore an understanding-oriented approach instead of a purely predictive one, it is crucial to remember that the ultimate objective remains the generation of effective strategies for preventing and detecting cyberbullying. The challenge lies in synergizing algorithmic precision with practical insights in a realm that has profound implications for the welfare of minors.

%%%%%%%%

% - Ciertas acciones que toma el jugador, diseñadas a propósito, parecen más representativas para identificar riesgo de cometer CB que los datos demográficos ("dime lo que haces y no lo que eres" "measuring risks without taking any")

% - Si se reduce el número de preguntas, se "exageran" los resultados obtenidos

% - Conclusión desde el punto de vista de policy making: convendría trabajar en las intervenciones para evitar ciertos comportamientos, a los que puede que no se le de mucha importancia (al menos, los propios menores)

%----------CONCLUSIONS---------
\section{Conclusions}
In this paper, we explore cyberbullying, a sensitive and pervasive issue affecting today's digital society, intending to improve the understanding of its underlying mechanisms. Using a novel instrument, a SG, we sought to observe and learn from minors' online behavior patterns in an engaging and non-invasive way. The choice of a SG served a dual purpose: first, it served as an interactive educational medium for minors, and second, it provided a dynamic platform for collecting data on online behavior. In this way, we reinforced the potential and validity of using SGs as practical social research tools.

We adopted a causality-based analytical perspective to navigate the intricate maze of data collected, specifically using BNs. This method allowed us to obtain more robust and interpretable results. Fundamentally, it facilitated a clear demonstration of the working model and underlying assumptions, thus fostering open, interpretable, and debatable science. This clarity could spark broader debate, stimulate further research, and ensure reproducibility, all essential in the quest for scientific advancement.

In addition to the gameplay data, we collected demographic information and validated psychological questionnaires to understand the participants better. Interestingly, our analysis revealed that serious play outperformed these traditional data collection methods regarding information. The immersive and interactive nature of the game provided more nuanced information about minors' online behavior, suggesting its potential for social science research.

However, it is essential to interpret these results with caution, considering the limited sample size and the inherent noise often found in social science and video game data. As we progress in this line of research, expanding the sample size and refining our analytic tools will be crucial steps in further affirming the value of SGs and causality-based analytics in studying complex social issues such as cyberbullying.

\section*{Acknowledgement}
This work has received funding from the European
Union’s Horizon 2020 research and innovation programme
under grant agreement No 882828. The authors would like to
thank all the partners within the consortium for the fruitful
collaboration and discussion. The sole responsibility for the
content of this document lies with the authors and in no way
reflects the views of the European Union.
This work has been partially supported by Grant PID2022-140217NB-I00 funded by MCIN/AEI/ 10.13039/501100011033.

%\clearpage

%\bibliographystyle{cas-model2-names}
\bibliographystyle{ieeetr}
%\bibliographystyle{plain}
\bibliography{cas-dc-template.bib}

\begin{thebibliography}{10}

\bibitem{UNICEFchildren2017}
UNICEF, ``{The State of the World's Children 2017: Children in a Digital
  World},'' tech. rep., UNICEF Division of Communication, 2017.
\newblock {ISBN: 978-92-806-4930-7}.

\bibitem{EUkidsOnline2020}
D.~Smahel, H.~Machackova, G.~Mascheroni, L.~Dedkova, E.~Staksrud,
  K.~{\'O}lafsson, S.~Livingstone, and U.~Hasebrink, ``{EU Kids Online 2020:
  Survey results from 19 countries},'' tech. rep., {EU Kids Online}, 2020.
\newblock {ISSN: 2045-256X}.

\bibitem{ChidrenRisk_Covid2021}
{European Commission}, J.~R. Centre, B.~Lobe, A.~Velicu, E.~Staksrud,
  S.~Chaudron, and R.~Di~Gioia, {\em How children (10-18) experienced online
  risks during the COVID-19 lockdown : Spring 2020 : key findings from
  surveying families in 11 European countries}.
\newblock Publications Office of the European Union, 2021.

\bibitem{abt1987serious}
C.~C. Abt, {\em Serious games}.
\newblock University press of America, 1987.

\bibitem{RAYUELA}
G.~L{\'o}pez, N.~Bueno, M.~Castro, M.~Reneses, J.~P{\'e}rez, M.~Riberas,
  M.~{\'A}lvarez-Campana, M.~Vega-Barbas, S.~Solera-Cotanilla, L.~Bastida, {\em
  et~al.}, ``{The H2020 project RAYUELA: A fun way to fight cybercrime},'' {\em
  Jornadas Nacionales de Investigación en Ciberseguridad - JNIC}, 2021.

\bibitem{gamesScience2023}
B.~Long, J.~Simson, A.~Bux{\'{o}}-Lugo, D.~G. Watson, and S.~A. Mehr, ``How
  games can make behavioural science better,'' {\em Nature}, vol.~613,
  pp.~433--436, Jan. 2023.

\bibitem{PerezSeriousGames}
J.~Pérez, M.~Castro, and G.~López, ``Serious games and ai: Challenges and
  opportunities for computational social science,'' {\em IEEE Access}, vol.~11,
  pp.~62051--62061, 2023.

\bibitem{Navigation2022}
A.~Coutrot, E.~Manley, S.~Goodroe, C.~Gahnstrom, G.~Filomena, D.~Yesiltepe,
  R.~C. Dalton, J.~M. Wiener, C.~H\"{o}lscher, M.~Hornberger, and H.~J. Spiers,
  ``Entropy of city street networks linked to future spatial navigation
  ability,'' {\em Nature}, vol.~604, pp.~104--110, Mar. 2022.

\bibitem{English2018}
J.~K. Hartshorne, J.~B. Tenenbaum, and S.~Pinker, ``A critical period for
  second language acquisition: Evidence from 2/3 million english speakers,''
  {\em Cognition}, vol.~177, pp.~263--277, Aug. 2018.

\bibitem{awad_moral_2018}
E.~Awad, S.~Dsouza, R.~Kim, J.~Schulz, J.~Henrich, A.~Shariff, J.-F. Bonnefon,
  and I.~Rah\-wan, ``The {Moral} {Machine} experiment,'' {\em Nature},
  vol.~563, pp.~59--64, Nov. 2018.

\bibitem{rahal2022rise}
C.~Rahal, M.~Verhagen, and D.~Kirk, ``The rise of machine learning in the
  academic social sciences,'' {\em AI \& SOCIETY}, pp.~1--3, 2022.

\bibitem{pearl2009causality}
J.~Pearl, {\em Causality}.
\newblock Cambridge university press, 2009.

\bibitem{Pearl_Do_Calculus}
J.~PEARL, ``{Causal diagrams for empirical research},'' {\em Biometrika},
  vol.~82, pp.~669--688, 12 1995.

\bibitem{rosenberg1965rosenberg}
M.~Rosenberg, ``Rosenberg self-esteem scale (rse),'' {\em Acceptance and
  commitment therapy. Measures package}, vol.~61, no.~52, p.~18, 1965.

\bibitem{SocialSupport}
G.~D. Zimet, N.~W. Dahlem, S.~G. Zimet, and G.~K. Farley, ``The
  multidimensional scale of perceived social support,'' {\em Journal of
  Personality Assessment}, vol.~52, no.~1, pp.~30--41, 1988.

\bibitem{CB_Questionnaire}
A.~Brighi, R.~Ortega, J.~Pyzalski, H.~Scheithauer, P.~K. Smith,
  H.~Tsormpatzoudis, H.~Tsorbatzoudis, and et~al., ``European cyberbullying
  intervention project questionnaire,'' 2012.

\bibitem{koiter2006visualizing}
J.~R. Koiter, ``Visualizing inference in bayesian networks,'' {M.Sc. thesis},
  Faculty of Electrical Engineering, Mathematics, and Computer Science,
  Department of Man-Machine Interaction, Delft University of Technology, 2006.

\bibitem{JSD}
D.~Endres and J.~Schindelin, ``A new metric for probability distributions,''
  {\em IEEE Transactions on Information Theory}, vol.~49, no.~7,
  pp.~1858--1860, 2003.

\bibitem{yannakakis2018artificial}
G.~N. Yannakakis and J.~Togelius, {\em \emph{Artificial intelligence and
  games}}, vol.~2.
\newblock Springer, 2018.

\bibitem{Jensen2007}
F.~V. Jensen and T.~D. Nielsen, {\em Bayesian Networks and Decision Graphs}.
\newblock Springer New York, 2007.

\bibitem{jeffreys1998theory}
H.~Jeffreys, {\em The theory of probability}.
\newblock OuP Oxford, 1998.

\bibitem{Spearman}
C.~Spearman, ``The proof and measurement of association between two things,''
  {\em The American Journal of Psychology}, vol.~15, no.~1, pp.~72--101, 1904.

\end{thebibliography}

\clearpage

\appendix
\section{Appendix: Game Decisions Transcript}
\label{appendix}

Note: The RAYUELA video game covers several cyber\-crimes affecting minors. Only Adventures 1 and 3 deal with cyberbullying, so these were the ones used in this paper.
%%%%%%%%%%%%%%%%%%
%%%%%%%%%%%%%%%%%%
%%%%%%%%%%%%%%%%%%
\subsection*{Adventure 1}

%%%%%%%%%%%%%%%%%%
\textbf{Question 1: Photo Sharing}

[Talking to Matthew after taking a selfie.]

\textit{Now we only have to share and tag the photos. Jane, do you want to share them, or do you prefer me to do it?}

\begin{itemize}[label=$\square$]
    \setlength\itemsep{0cm}
    \item \textit{I will do it.}
    \item \textit{You can do it.}
\end{itemize}

%%%%%%%%%%%%%%%%%%
\textbf{Question 2: Sociable}

[Talking to Robert after sharing the selfie. Dialog depends on the previous answers.]

\textit{It seems you like to upload many photos and share stuff on your social network.}

\begin{itemize}[label=$\square$]
    \setlength\itemsep{0cm}
    \item \textit{I would say I am sociable.}
    \item \textit{I consider myself kind of shy.}
\end{itemize}

%%%%%%%%%%%%%%%%%%
\textbf{Question 3: Matthew Meme}

[After receiving a message from Patty with the meme about Matthew.]

\begin{itemize}[label=$\square$]
    \setlength\itemsep{0cm}
    \item \textit{Hehe, it's funny, I will share the meme.}
    \item \textit{I am not going to share it.}
    \item \textit{I won't share the meme and I'm going to try to end up with this.}
\end{itemize}

%%%%%%%%%%%%%%%%%%
%%%%%%%%%%%%%%%%%%
%%%%%%%%%%%%%%%%%%
\subsection*{Adventure 2}

%%%%%%%%%%%%%%%%%%
\textbf{Question 1: Registration Name}

[Creating a new profile on a social network. The user has to select a profile name.]

\begin{itemize}[label=$\square$]
    \setlength\itemsep{0cm}
    \item \textit{My name and my year of birth.}
    \item \textit{My name and surname.}
    \item \textit{My favourite music band name.}
    \item \textit{Other famous/TV/Book character I like.}
\end{itemize}

%%%%%%%%%%%%%%%%%%
\textbf{Question 2: Registration Password}

[The user has to select a profile password.]

\begin{itemize}[label=$\square$]
    \setlength\itemsep{0cm}
    \item \textit{I don’t have time for this; I will leave the default password.}
    \item \textit{My name and surname.}
    \item \textit{I'll use the same password I have on other websites, so it's easier to remember.}
    \item \textit{I am going to set a strong password, even if I have to invest some more time.}
\end{itemize}

%%%%%%%%%%%%%%%%%%
\textbf{Question 3: Registration Profile Type}

[The user has to select a profile type.]

\begin{itemize}[label=$\square$]
    \setlength\itemsep{0cm}
    \item Public profile.
    \item Private profile.
\end{itemize}

%%%%%%%%%%%%%%%%%%
\textbf{Question 4: Registration Profile Place}

[The user has to select a profile place.]

\begin{itemize}[label=$\square$]
    \setlength\itemsep{0cm}
    \item \textit{The name of the city and neighbourhood where I live.}
    \item \textit{The name of my school and country.}
    \item \textit{Something fantastic, as "I am in the clouds", "In the moon" or "Too far away from X".}
    \item \textit{Leave empty.}
\end{itemize}

%%%%%%%%%%%%%%%%%%
\textbf{Question 5: Registration Profile Photo}

[The user has to select a profile photo.]

\begin{itemize}[label=$\square$]
    \setlength\itemsep{0cm}
    \item \textit{A photo of just me.}
    \item \textit{A photo of me and some friends.}
    \item \textit{A photo from the Internet, in which I do not appear.}
\end{itemize}

%%%%%%%%%%%%%%%%%%
\textbf{Question 6: Comment Patty Post}

[Seeing a post from Patty on the social network.]

\begin{itemize}[label=$\square$]
    \setlength\itemsep{0cm}
    \item \textit{Good one!}
    \item \textit{They are awesome.}
    \item \textit{I don't like them, it's so childish.}
    \item \textit{Don't send any comment. I'm sure she won't pay any attention to it.}
\end{itemize}

%%%%%%%%%%%%%%%%%%
\textbf{Question 7: Use PC}

[Using the club's PC after accepting a friend request from a photographer on the social network.]

\begin{itemize}[label=$\square$]
    \setlength\itemsep{0cm}
    \item \textit{View messages.}
    \item \textit{Check photographer's profile.}
\end{itemize}

%%%%%%%%%%%%%%%%%%
\textbf{Question 8: Friend Request}

[Using the club's PC.]

\begin{itemize}[label=$\square$]
    \setlength\itemsep{0cm}
    \item \textit{Accept friend request.}
    \item \textit{Reject friend request.}
    \item \textit{Check photographer's profile.}
\end{itemize}

%%%%%%%%%%%%%%%%%%
\textbf{Question 9: Send Photos}

[Checking messages after accepting the friend request.]

\begin{itemize}[label=$\square$]
    \setlength\itemsep{0cm}
    \item \textit{Send photos.}
    \item \textit{Not send photos.}
\end{itemize}

%%%%%%%%%%%%%%%%%%
\textbf{Question 10: More Photos}

[Checking messages after sending swimsuit photos.]

\begin{itemize}[label=$\square$]
    \setlength\itemsep{0cm}
    \item \textit{Send naked photos.}
    \item \textit{Do not send naked photos.}
\end{itemize}

%%%%%%%%%%%%%%%%%%
\textbf{Question 11: More \& More}

[Checking messages after sending naked photos.]

\begin{itemize}[label=$\square$]
    \setlength\itemsep{0cm}
    \item \textit{Send more naked photos.}
    \item \textit{Reject the request and inform Mary.}
\end{itemize}

%%%%%%%%%%%%%%%%%%
\textbf{Question 12: Ask Help}

\begin{itemize}[label=$\square$]
    \setlength\itemsep{0cm}
    \item \textit{Ask for help to Mary.}
    \item \textit{Say nothing.}
\end{itemize}

%%%%%%%%%%%%%%%%%%
\textbf{Question 13: Close Case}

\begin{itemize}[label=$\square$]
    \setlength\itemsep{0cm}
    \item \textit{Ok, I'll check the profile.}
    \item \textit{No, I won't check the profile.}
\end{itemize}

%%%%%%%%%%%%%%%%%%
\textbf{Question 14: Tell Parents}

[At the end of the scene. Previous dialog depends on player decisions.]

\textit{We should report that profile to the social network. Besides, you should also tell your parents about it and see if we need to talk with the police.}

\begin{itemize}[label=$\square$]
    \setlength\itemsep{0cm}
    \item \textit{I don’t know, Mary, communication with them is not very easy. Lately they get angry about anything and we always end up shouting.}
    \item \textit{I'm ashamed! I don't want to tell them something like that. It's better if I try to solve it on my own.}
    \item \textit{Yeah, you’re probably right. I'm a bit embarrassed but I'll give it a try.}
\end{itemize}

%%%%%%%%%%%%%%%%%%
\textbf{Question 15: Block Profile}

\begin{itemize}[label=$\square$]
    \setlength\itemsep{0cm}
    \item \textit{Block the profile.}
    \item \textit{Do not block the profile.}
\end{itemize}

%%%%%%%%%%%%%%%%%%
%%%%%%%%%%%%%%%%%%
%%%%%%%%%%%%%%%%%%
\subsection*{Adventure 3}

%%%%%%%%%%%%%%%%%%
\textbf{Question 1: Pirated Content}

[Playing video games in your room.]

\textit{I know of some sites that pirate the content and then you can download the update for free.}

\begin{itemize}[label=$\square$]
    \setlength\itemsep{0cm}
    \item \textit{I will download the pirated update through a website.}
    \item \textit{I will wait until I have money or until my parents give me the money to buy the new expansion.}
\end{itemize}

%%%%%%%%%%%%%%%%%%
\textbf{Question 2: Pol or Paula}

[Playing video games in your room. Your friends joke about Pol's appearance.]

\begin{itemize}[label=$\square$]
    \setlength\itemsep{0cm}
    \item \textit{I don’t like this kind of jokes.}
    \item \textit{That’s funny.}
    \item \textit{Say nothing.}
\end{itemize}

%%%%%%%%%%%%%%%%%%
\textbf{Question 3: Time Overrun}

[Playing video games in your room. A warning pops up about the number of hours you have been online]

\begin{itemize}[label=$\square$]
    \setlength\itemsep{0cm}
    \item \textit{4 hours are not that much. So, I can keep chatting a bit longer.}
    \item \textit{It's time to stop and disconnect for a while, although I might miss some juicy gossiping.}
\end{itemize}

%%%%%%%%%%%%%%%%%%
\textbf{Question 4: Pol Bullied}

\textit{I heard that some guys are messing with Paul. I am worried that he may be bullied. What do you think?}

\begin{itemize}[label=$\square$]
    \setlength\itemsep{0cm}
    \item \textit{They are just having fun; I would not call that bullying.}
    \item \textit{I think it’s not right... but calling that bullying is a bit of a stretch.}
    \item \textit{I think that’s unacceptable; we should do something about it.}
\end{itemize}

%%%%%%%%%%%%%%%%%%
\textbf{Question 5: Remind Matthew}

[Your friends start messing with Pol.]

\begin{itemize}[label=$\square$]
    \setlength\itemsep{0cm}
    \item \textit{Yes, I had a similar bad experience... I don't like being picked on.}
    \item \textit{No, it has never really happened to me, to my knowledge.}
    \item \textit{Yes, it was me who messed with someone else... but it was not such a big deal.}
\end{itemize}

%%%%%%%%%%%%%%%%%%
\textbf{Question 6: Talk to Pol}

\textit{So, shall we talk to Pol to see how he is?}

\begin{itemize}[label=$\square$]
    \setlength\itemsep{0cm}
    \item \textit{It is better to let him be.}
    \item \textit{Of course, we should try to help.}
\end{itemize}

%%%%%%%%%%%%%%%%%%
\textbf{Question 7: How to Help Pol}

\textit{We can help you. You are not alone in this. I think...}

\begin{itemize}[label=$\square$]
    \setlength\itemsep{0cm}
    \item \textit{We should  go to tell the teacher, he should know what to do.}
    \item \textit{We should report the comments to the social network, so that it doesn't happen again.}
    \item \textit{We should not report it, because I don’t want to get picked on for being a snitch…}
    \item \textit{We should not report it as reporting is usually useless.}
\end{itemize}

%%%%%%%%%%%%%%%%%%
%%%%%%%%%%%%%%%%%%
%%%%%%%%%%%%%%%%%%
\subsection*{Adventure 4}

%%%%%%%%%%%%%%%%%%
\textbf{Question 1: Phising Email}

[You receive an email indicating that your social network account has been compromised.]

\begin{itemize}[label=$\square$]
    \setlength\itemsep{0cm}
    \item \textit{Is my account in danger?! I must act quickly before I lose it. I will follow the instructions in the email.}
    \item \textit{I find it suspicious...I’ll better go straight to my social network profile’s security settings and change my password there.}
\end{itemize}

%%%%%%%%%%%%%%%%%%
\textbf{Question 2: New Password}

[You proceed to change the password of your account.]

\begin{itemize}[label=$\square$]
    \setlength\itemsep{0cm}
    \item (Password = Name123) \textit{I'm going to leave a password very similar to the one I had before. Otherwise, I'll forget it...}
    \item (Password = Football10) \textit{I'm going to make a password with some hobbies or things I like in it. So, I won't forget it!}
    \item (Password = Ax/2oP3\%nY6) \textit{I'm going to make my password difficult and long. It is more challenging this way, but much safer.}
\end{itemize}

%%%%%%%%%%%%%%%%%%
\textbf{Question 3: My Account Stolen}

[If your account has been phished.]

\begin{itemize}[label=$\square$]
    \setlength\itemsep{0cm}
    \item \textit{It does not seem to be that worrying, it’s just a social network account. We don’t need to tell this to anyone. I can create another account after all.}
    \item \textit{It is important to tell someone or report it, since the account contains personal information. It is a crime!}
\end{itemize}

%%%%%%%%%%%%%%%%%%
\textbf{Question 4: Other Account Stolen}

[If John has not been phished.]

\begin{itemize}[label=$\square$]
    \setlength\itemsep{0cm}
    \item \textit{It does not seem to be that worrying, it’s just a social network account. We don’t need to tell this to anyone. He can create another account after all.}
    \item \textit{It is important to tell someone or report it, since his account contains personal information. It is a crime!}
\end{itemize}

%%%%%%%%%%%%%%%%%%
%%%%%%%%%%%%%%%%%%
%%%%%%%%%%%%%%%%%%
\subsection*{Adventure 5}

%%%%%%%%%%%%%%%%%%
\textbf{Question 1: Secret Relationship}

[You and your friends are commenting that Sheila has a new romantic relationship that is distancing her from her friends and you are worried.]

\begin{itemize}[label=$\square$]
    \setlength\itemsep{0cm}
    \item \textit{Love is love, and everyone experiences it in a different way. If she needed help, she would have asked for it, wouldn't she?}
    \item \textit{Sounds a bit creepy to me, have you tried looking at her social media?}
\end{itemize}

%%%%%%%%%%%%%%%%%%
\textbf{Question 2: Biology Paper}

[You must meet to do a biology assignment and indicate your preference to meet online or in person.]

\begin{itemize}[label=$\square$]
    \setlength\itemsep{0cm}
    \item \textit{Meet at the library this afternoon, so we can go to the cafeteria if we finish earlier.}
    \item \textit{Do it by video-conference this afternoon, so we can be more comfortable at home.}
\end{itemize}

%%%%%%%%%%%%%%%%%%
\textbf{Question 3: Talk Sheila}

[John sees Sheila, who is leaning against the wall with her eyes fixed on her mobile phone.]

\begin{itemize}[label=$\square$]
    \setlength\itemsep{0cm}
    \item \textit{I will talk to her and let her know she can trust me if she has any problem.}
    \item \textit{I should text Mary since she is closer to Sheila.}
\end{itemize}

%%%%%%%%%%%%%%%%%%
%%%%%%%%%%%%%%%%%%
%%%%%%%%%%%%%%%%%%
\subsection*{Adventure 6}

%%%%%%%%%%%%%%%%%%
\textbf{Question 1: Migrant News Check}

[When investigating a news item that appears to be false, you must decide which things seem most relevant to verify the information.]

\begin{itemize}[label=$\square$]
    \setlength\itemsep{0cm}
    \item \textit{How professional the web page looks like (style, images, design, etc).}
    \item \textit{The source itself: is it a known newspaper/website or is it an unknown site?}
    \item \textit{If the information looks accurate, for instance with enough numbers and statistics.}
    \item \textit{Search on the Internet to contrast the information.}
\end{itemize}

%%%%%%%%%%%%%%%%%%
\textbf{Question 2: Web Page Looks Like}

[Reviewing the website.]

\begin{itemize}[label=$\square$]
    \setlength\itemsep{0cm}
    \item \textit{It seems it is true. Definitely not a fake page.}
    \item \textit{It looks quite professional, but does it mean it’s not fake? We should try other options.}
\end{itemize}

%%%%%%%%%%%%%%%%%%
\textbf{Question 3: The Source}

[Reviewing the source.]

\begin{itemize}[label=$\square$]
    \setlength\itemsep{0cm}
    \item \textit{It is a known newspaper, at least I’ve seen it quite a lot on Social Networks. I would say is not fake.}
    \item \textit{Even though it is a kind of famous newspaper, it could contain fake information. We should try other options.}
\end{itemize}

%%%%%%%%%%%%%%%%%%
\textbf{Question 4: Information Looks Accurate}

[Reviewing if the information looks accurate.]

\begin{itemize}[label=$\square$]
    \setlength\itemsep{0cm}
    \item \textit{Ok, they are displaying a big amount of data, and look at the graph as it rises. It looks pretty accurate.}
    \item \textit{Ok, there are a lot of numbers and graphs, but that does not mean that the data is correct. Data can also be falsified.}
\end{itemize}

%%%%%%%%%%%%%%%%%%
\textbf{Question 5: Replay post}

\textit{Ok, so first of all, we should report the content to the social network, and we should probably reply with this information, right?}

\begin{itemize}[label=$\square$]
    \setlength\itemsep{0cm}
    \item \textit{It is not worth answering. Don't feed the troll!}
    \item \textit{Yes, let’s add the link to the anti-hoaxes website.}
\end{itemize}

%%%%%%%%%%%%%%%%%%
\textbf{Question 6: Regarding Charles}

\textit{What should we do with Charles?}

\begin{itemize}[label=$\square$]
    \setlength\itemsep{0cm}
    \item \textit{It is a basket case, there is little we can do.}
    \item \textit{We should try to talk to him.}
\end{itemize}

% \section{Exploratory Data Analysis}
% \label{appendixEDA}

\end{document}